\documentclass[a4paper]{article}

\usepackage[utf8]{inputenc}
\usepackage{authblk}
\usepackage{amsmath}
\usepackage{amsthm}
\usepackage{bbold}
\usepackage{enumitem}
\usepackage[autosize]{dot2texi}
\usepackage{tikz}
\usetikzlibrary{shapes,arrows}
\usepackage{subcaption}
\usepackage{xcolor}
\usepackage[toc]{appendix}
\usepackage{hyperref}
\usepackage{graphicx}

\title{More Clustering Quality Metrics for ABCDE}
\author[1]{Stephan van Staden}
\affil[1]{Google Switzerland GmbH}
\date{September 2024}

\newcommand{\Base}{\mathit{Base}}
\newcommand{\Exp}{\mathit{Exp}}
\newcommand{\Ideal}{\mathit{Ideal}}
\newcommand{\weight}{\mathit{weight}}
\newcommand{\JaccardDistance}{\mathit{JaccardDistance}}

\newcommand{\SplitRate}{\mathit{SplitRate}}
\newcommand{\MergeRate}{\mathit{MergeRate}}
\newcommand{\Precision}{\mathit{Precision}}
\newcommand{\Recall}{\mathit{Recall}}

\newcommand{\GoodSplitRate}{\mathit{GoodSplitRate}}
\newcommand{\BadSplitRate}{\mathit{BadSplitRate}}
\newcommand{\GoodMergeRate}{\mathit{GoodMergeRate}}
\newcommand{\BadMergeRate}{\mathit{BadMergeRate}}
\newcommand{\GoodBadSplitMergeRate}{\mathit{(Good|Bad)(Split|Merge)Rate}}
\newcommand{\DeltaPrecision}{\Delta\mathit{Precision}}
\newcommand{\DeltaRecall}{\Delta\mathit{Recall}}

\newcommand{\indicate}{\mathbb{1}}

\newcommand{\match}{\equiv}

\newcommand{\IQ}{\mathit{IQ}}
\newcommand{\GoodSplitWeight}{\mathrm{GoodSplitWeight}}
\newcommand{\BadSplitWeight}{\mathrm{BadSplitWeight}}
\newcommand{\GoodMergeWeight}{\mathrm{GoodMergeWeight}}
\newcommand{\BadMergeWeight}{\mathrm{BadMergeWeight}}
\newcommand{\GoodStableWeight}{\mathrm{GoodStableWeight}}
\newcommand{\BadStableWeight}{\mathrm{BadStableWeight}}
\newcommand{\MissingWeight}{\mathrm{MissingWeight}}
\newcommand{\StableWeight}{\mathrm{StableWeight}}
\newcommand{\AffectedItems}{\mathit{AffectedItems}}
\newcommand{\UnaffectedItems}{\mathit{UnaffectedItems}}
\newcommand{\AffectedWeightFraction}{\mathit{AffectedWeightFraction}}
\newcommand{\JaccardDistanceToIdealImprovement}{\mathit{JaccardDistanceToIdealImprovement}}
\newcommand{\PerfectPrecision}{\mathit{PerfectPrecision}}
\newcommand{\PerfectRecall}{\mathit{PerfectRecall}}
\newcommand{\MinimumRecall}{\mathit{MinimumRecall}}

\begin{document}

\maketitle

\begin{abstract}
ABCDE is a technique for evaluating clusterings of very large populations of items. Given two clusterings, namely a Baseline clustering and an Experiment clustering, ABCDE can characterize their differences with impact and quality metrics, and thus help to determine which clustering to prefer. We previously described the basic quality metrics of ABCDE, namely the $\GoodSplitRate$, $\BadSplitRate$, $\GoodMergeRate$, $\BadMergeRate$ and $\DeltaPrecision$, and how to estimate them on the basis of human judgements. This paper extends that treatment with more quality metrics. It describes a technique that aims to characterize the $\DeltaRecall$ of the clustering change. It introduces a new metric, called $\IQ$, to characterize the degree to which the clustering diff translates into an improvement in the quality. Ideally, a large diff would improve the quality by a large amount. Finally, this paper mentions ways to characterize the absolute $\Precision$ and $\Recall$ of a single clustering with ABCDE.
\end{abstract}

{\bf Keywords:} Clustering evaluation, Clustering metrics, Clustering quality, ABCDE, Delta Recall, IQ, Absolute Precision, Absolute Recall

\section{Introduction}

ABCDE, which stands for `Application-Based Cluster Diff Evals'~\cite{vanstadengrubb2024abcde}, is a technique for evaluating clusterings of very large populations of items. Given two clusterings, namely a Baseline clustering (henceforth referred to as $\Base$) and an Experiment clustering (henceforth referred to as $\Exp$), ABCDE can characterize their differences with impact and quality metrics, and thus help to determine which clustering to prefer. The original paper~\cite{vanstadengrubb2024abcde} described the basic metrics of ABCDE, namely:
\begin{itemize}
    \item Impact metrics \\ $\SplitRate$, $\MergeRate$, $\JaccardDistance$. \\ They can be computed exactly.
    \item Quality metrics \\ $\GoodSplitRate$, $\BadSplitRate$, $\GoodMergeRate$, $\BadMergeRate$, $\DeltaPrecision$. \\ They are estimated on the basis of human judgements about pairs of items. The sampling of the pairs, and the estimation of the metrics from the judgements, are rigorously stipulated by ABCDE.
\end{itemize}

The basic metrics of ABCDE already provide a lot of information, but they omit some interesting aspects. The goal of this paper is to fill the gaps with additional metrics. As we will see, these metrics either characterize or involve the clustering quality, and some of them make simplifying assumptions and employ approximate reasoning to make the measurement task tractable.

The first topic is about characterizing the $\DeltaRecall$ of the clustering change. As mentioned in~\cite{vanstaden2024pointwise}, the $\Precision$ and the $\Recall$ metrics go hand in hand: $\Precision$ characterizes the homogeneity of clusters, while $\Recall$ characterizes their completeness. One can trivially maximize $\Precision$ by putting each item in a cluster of its own, but then $\Recall$ will suffer. Likewise, putting all items in one single huge cluster will maximize $\Recall$ at the expense of $\Precision$. Since ABCDE provides a statistical estimate of $\DeltaPrecision$, it is natural to consider also $\DeltaRecall$.

Obtaining an estimate for $\DeltaRecall$ is non-trivial in the context of diff evals for very large clusterings. Given an arbitrary item, the search space in which to look for other items that are equivalent to it is huge, which makes it similar to looking for a few needles in a haystack. Moreover, a diff eval can afford only a very limited number of human judgements because of cost and latency reasons: developers do not want to wait for weeks for a $\DeltaRecall$ estimate of an experiment, and the associated cost to obtain many judgements for many experiments would be enormous.

There are several ways to deal with this dilemma. For example, one might decide that putting a number on $\DeltaRecall$ is not so important for the application at hand, and consider only concrete examples of $\Recall$ wins (good merges) and $\Recall$ losses (bad splits). Or one might obtain a value for $\DeltaRecall$ from an evaluation with respect to some ground truth clustering, as described in~\cite{vanstaden2024pointwise}. The ground truth clustering can be constructed with $\Recall$ measurements in mind\footnote{For example, select a few hundred items, none of which are equivalent to each other, and for each one use some retrieval mechanism (e.g. human searches based on keywords, or nearest neighbors in an embedding space) to find say 1000 other items that are potentially equivalent to it, and ask humans which of them are indeed equivalent. Each original item together with the other items that are truly equivalent to it then form a ground truth cluster. Assessing the $\Recall$ of $\Exp$ and $\Base$ with respect to this ground truth clustering and deducting the results measures the $\DeltaRecall$ with respect to the items mentioned in the ground truth clustering.}. Since the ground truth clustering can be persisted and reused many times to evaluate new experimental clusterings, the long time it took to obtain its human judgements is soon forgotten and their amortized cost is low. But the set of items that are considered in the evaluation is then small and fixed, which is exactly what ABCDE aims to overcome.

The technique described in this paper equips ABCDE with an approximate value for $\DeltaRecall$. More precisely, it uses sophisticated back-of-the-envelope reasoning to express $\DeltaRecall$ as a function of the baseline quality of the items that experienced change. The reasoning leverages the rich information supplied by the basic ABCDE metrics. It also uses some auxiliary measurements about the clusterings that are easy to compute and do not require human judgement.

The second topic of this paper is about characterizing the degree to which the clustering diff translates into an improvement in the quality of the clustering. Ideally, a large clustering diff would improve the quality by a large amount. A large clustering diff that has neutral or negative quality is usually undesirable, unless it comes with big improvements in efficiency, infrastructure costs or maintainability. A new ABCDE metric, called $\IQ$, expresses the relationship between the Impact and the Quality of the clustering change. A value of $\IQ = 100\%$ means that everything in the diff serves to improve the quality of the clustering, $\IQ = 0\%$ means that the diff has neutral quality, and $\IQ = -100\%$ means that everything in the diff worsens the quality.

The third and last topic of this paper is about characterizing the absolute $\Precision$ and $\Recall$ of a single clustering, i.e. a clustering snapshot. We normally use ABCDE to characterize the relative differences between two clusterings with delta metrics. Yet it can be informative to know the absolute metrics of the clustering snapshot that is currently deployed/used. Tracking absolute metrics over time can be useful to see progress, identify regressions, and set priorities for further development. As we will see, ABCDE can help to understand the absolute quality of a single clustering snapshot. Doing so typically needs more human judgements compared to the delta quality metrics of day-to-day experiments, but it can be feasible every now and then. In the intervals between such measurements, we can use the delta metrics of the accepted clustering changes to get an idea about the quality of the current snapshot.

\section{Preliminaries}

Given a finite set of items $T$ and an equivalence relation $\match$ (i.e. a binary relation that is reflexive, symmetric and transitive), a \textit{cluster} is an equivalence class of $T$ with respect to $\match$, and a \textit{clustering} is the set of all clusters, i.e. a partitioning of $T$ into its equivalence classes.

In practical applications, the set of items $T$ can be very large and the ideal equivalence relation is not fully known. Humans can consider a pair of items and say whether they are truly equivalent or not, but since that does not scale to billions of items, we have only very sparse information about the ideal relation. The main job of a clustering algorithm in such a setting is to approximate the ideal equivalence relation. This is typically done by 1) deciding which items might be related (also called `blocking'), and 2) deciding which of these are equivalent according to a computable function that imitates the human judgements. The design space of clustering algorithms is consequently huge, and the goal of ABCDE is to evaluate the clustering results to help determine which algorithm and configuration to prefer in practice.

So in the context of ABCDE, we have an approximated equivalence relation for $\Base$, and an approximated equivalence relation for $\Exp$. In fact $\Base$ and $\Exp$ might have entirely different sets of clusters, so the approximations do not have to be related in any way. The rest of this paper will not refer to approximated equivalence relations anymore. Instead, it will simply consider clusters, i.e. sets of items that are considered equivalent by a given clustering, and it will use $\match$ to denote the ideal equivalence relation. Any given pair of clusterings can be restricted to the items they have in common, as described in~\cite{vanstaden2024pointwise}. The rest of the paper assumes that $\Base$ and $\Exp$ are two clusterings of the same population of items $T$. It uses the definitions and notation of~\cite{vanstadengrubb2024abcde}, and extends that when needed.

\section{Reasoning about $\DeltaRecall$}

Let $\Ideal(i)$ denote the conceptually ideal cluster that contains the item $i$, i.e. the set of all items that are truly equivalent to $i$:
\begin{align}
\Ideal(i) = \{i^\prime \in T | i \match i^\prime\}
\end{align}

The $\Recall$ from the perspective of a given item $i$ of a clustering $C$ is defined as:
\begin{align}
\Recall_\mathit{C}(i) = \frac{\weight(\mathit{C}(i) \cap \Ideal(i))}{\weight(\Ideal(i))}
\end{align}
It therefore characterizes the degree to which the cluster $\mathit{C}(i)$ is complete.

$\Recall_\mathit{C}$ is a pointwise clustering metric~\cite{vanstaden2024pointwise} that is lifted to sets by using expected values, i.e. weighted averages. So for a set of items $I$, we have:
\begin{align}
\Recall_\mathit{C}(I) = \frac{\sum_{i \in I} weight(i) \cdot \Recall_\mathit{C}(i)}{\weight(I)}
\end{align}
where $\weight(I) = \sum_{i \in I} \weight(i)$.

In the context of ABCDE, we are primarily interested in the overall $\DeltaRecall$:
\begin{align}
\DeltaRecall(T) = \Recall_\Exp(T) - \Recall_\Base(T)
\end{align}
which is equal to the expected value of the per-item $\DeltaRecall$ over all items $T$, where
\begin{align}
\DeltaRecall(i) &= \frac{\weight(\mathit{\Exp}(i) \cap \Ideal(i)) - \weight(\mathit{\Base}(i) \cap \Ideal(i))}{\weight(\Ideal(i))} \\
&= \frac{1}{\weight(\Ideal(i))} \sum_{j \in \Base(i) \ominus \Exp(i)} u_j l_j \indicate(i \match j) \\
&~~~~\mathrm{if} \ j \in \Exp(i) \setminus \Base(i) :\ u_j = \weight(j),\ l_{j} = 1  \nonumber \\
&~~~~\mathrm{if} \ j \in \Base(i) \setminus \Exp(i) :\ u_j = \weight(j),\ l_{j} = -1  \nonumber
\end{align}
We can therefore express the overall $\DeltaRecall$ in terms of pairs of items:
\begin{align}
\DeltaRecall(T) &= \sum_{i \in T} \frac{\weight(i)}{\weight(T)} \DeltaRecall(i) \\
&= \sum_{i \in T} \sum_{j \in \Base(i) \ominus \Exp(i)} u_{ij} l_{ij} \indicate(i \match j) \\
\mathrm{if} \ j \in \Exp(i) \setminus \Base(i) :\ u_{ij} = \frac{\weight(i)}{\weight(T)}\frac{\weight(j)}{\weight(\Ideal(i))},\ l_{ij} = 1 ~~\span \nonumber  \\
\mathrm{if} \ j \in \Base(i) \setminus \Exp(i) :\ u_{ij} = \frac{\weight(i)}{\weight(T)}\frac{\weight(j)}{\weight(\Ideal(i))},\ l_{ij} = -1 \span \nonumber
\end{align}
The standard estimation procedure would be:
\begin{enumerate}
\item Sample pairs of items $(i, j)$, where $j \in \Base(i) \ominus \Exp(i)$, according to their weights $u_{ij}$.
\item Compute the average over the sample of $l_{ij} \indicate(i \match j)$.
\item Multiply that by $$\sum_{i \in T} \sum_{j \in \Base(i) \ominus \Exp(i)} u_{ij}$$ to obtain an estimate of $\DeltaRecall(T)$.
\end{enumerate}

While the definitions are straightforward, there is a catch in practice: it is intractable to know or even estimate $\weight(\Ideal(i))$ when the population $T$ contains billions of items and a huge number of ideal clusters. It is therefore intractable to obtain a statistical estimate of $\DeltaRecall(T)$. Even a valiant effort to look for a equivalent items will miss some and will find many that are not truly equivalent. The result is not very satisfying: it requires more human judgements than a diff eval can afford, and the result is still a biased underestimate.

The rest of this section takes an alternative approach. It describes a technique that aims to provide an approximate understanding of how a clustering change affects $\DeltaRecall$. The technique makes some simplifying assumptions to render the task tractable. It uses sophisticated back-of-the-envelope reasoning that leverages the rich information of the basic quality metrics of ABCDE. The reasoning does not require additional human judgements, which is great, but it provides only a rough understanding of the true overall $\DeltaRecall$ value of the clustering change.

We describe the technique by first looking at how it works on the level of individual items. This reasoning to compute $\DeltaRecall$ for an individual item is exact and parametric in $\Recall_\Base(i)$, and possibly also $\Precision_\Base(i)$.
Then we introduce the part that transfers the reasoning to the overall level, i.e. to the whole population of items $T$. That is approximate, because it involves some assumptions that are not true in general. The result should therefore be understood as a back-of-the-envelope estimate of how the clustering change affects the overall $\DeltaRecall$.

\subsection{Exact reasoning for individual items}

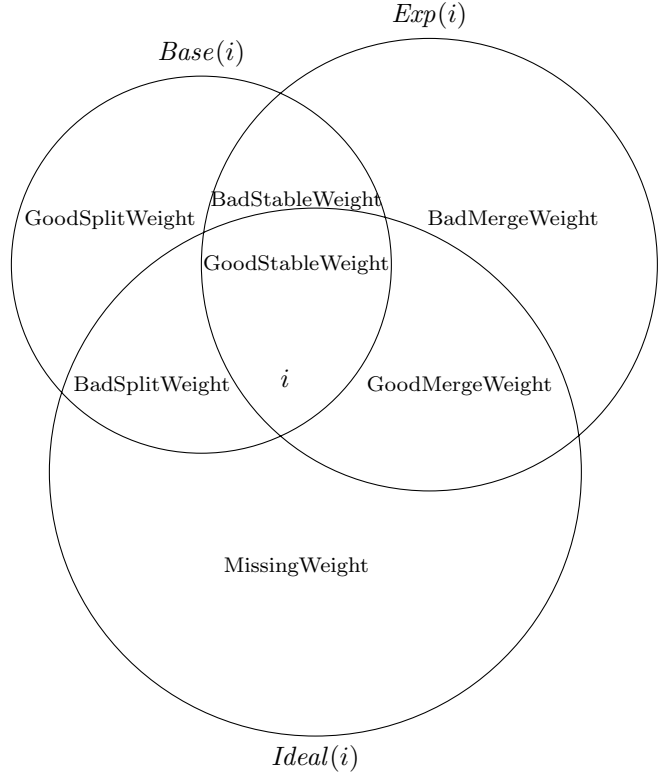
\begin{figure*}[t!]
\centering
\begin{tikzpicture}[fill=gray]
\draw (-1.5,0) circle (2.5)
      (-1.5,2.5)  node [text=black,above] {$\Base(i)$}
      (1.5,0) circle (3)
      (1.5,3)  node [text=black,above] {$\Exp(i)$}
      (-0.4,-1.5) node [text=black] {$i$}
      (0,-2.75) circle (3.5)
      (0,-6.25) node [text=black,below] {$\Ideal(i)$}
      (-0.25,0) node [text=black] {\footnotesize $\GoodStableWeight$}
      (-0.25,0.85) node [text=black] {\footnotesize $\BadStableWeight$}
      (-2.7,0.6) node [text=black] {\footnotesize $\GoodSplitWeight$}
      (-2.15,-1.6) node [text=black] {\footnotesize $\BadSplitWeight$}
      (2.6,0.6) node [text=black] {\footnotesize $\BadMergeWeight$}
      (1.9,-1.6) node [text=black] {\footnotesize $\GoodMergeWeight$}
      (-0.25,-4) node [text=black] {\footnotesize $\MissingWeight$}
      ;
\end{tikzpicture}
\caption{The clustering quality situation from the perspective of item $i$. The item $i$ is always in the intersection of $\Base(i)$ and $\Exp(i)$ and $\Ideal(i)$, which is never empty. $\Ideal(i)$ is the set of all items that are truly equivalent to $i$. Each area inside the Venn diagram is labeled with its weight. For example, $\GoodSplitWeight = \weight((\Base(i) \setminus \Exp(i)) \setminus \Ideal(i))$ is the weight of all items that were correctly split off from the perspective of $i$. Note that the items in $(\Base(i) \setminus \Exp(i)) \setminus \Ideal(i)$ are not equivalent to $i$, so removing them from the cluster of $i$ is good.}\label{base_exp_ideal_cluster_diagram}
\end{figure*}

Consider the clustering quality situation from the perspective of a given item $i$, as shown in Figure~\ref{base_exp_ideal_cluster_diagram}. It is not tractable to know all the members of $\Ideal(i)$ when $T$ has billions of items. In particular, the magnitude of $\MissingWeight$ in Figure~\ref{base_exp_ideal_cluster_diagram} is not known. So the reasoning here will consider how $\DeltaRecall(i)$ depends on the baseline $\Recall$ of $i$. It will express $\DeltaRecall(i)$ as a function of $\Recall_\Base(i)$, a number in the range $(0, 1]$ whose definition is taken from~\cite{vanstaden2024pointwise}:

\begin{align}
\Recall_\Base(i) &= \frac{\BadSplitWeight + \GoodStableWeight}{\weight(\Ideal(i))} \\
&= \frac{\BadSplitWeight + \GoodStableWeight}{\BadSplitWeight + \GoodStableWeight + \GoodMergeWeight + \MissingWeight}
\end{align}

The formula of the function can also use the ABCDE quality metrics of $i$, namely $\GoodSplitRate$, $\BadSplitRate$, $\GoodMergeRate$, $\BadMergeRate$ and $\DeltaPrecision$, as well as quantities that do not involve human judgements, for example $\weight(\Base(i))$ and $\weight(\Exp(i))$.

From the definitions of $\BadSplitRate$ and $\GoodMergeRate$ in~\cite{vanstadengrubb2024abcde}, we know the values of:
\begin{align}
\BadSplitWeight  &=  \BadSplitRate(i) \cdot \weight(\Base(i)) \\
\GoodMergeWeight &= \GoodMergeRate(i) \cdot \weight(\Exp(i))
\end{align}

So assuming we know the value of $\Recall_\Base(i)$, if we have a way to determine the value of $\GoodStableWeight$, then we would be able to determine $\weight(\Ideal(i))$ and hence also $\MissingWeight$:
\begin{align}
\weight(\Ideal(i)) = &\frac{\BadSplitWeight + \GoodStableWeight}{\Recall_\Base(i)} \\
\MissingWeight = &\frac{\BadSplitWeight + \GoodStableWeight}{\Recall_\Base(i)} \nonumber \\
& - \BadSplitWeight - \GoodStableWeight - \GoodMergeWeight \label{missing_weight_formula}
\end{align}

Once we know $\MissingWeight$, we can easily determine $\Recall_\Exp(i)$ and hence $\DeltaRecall(i) = \Recall_\Exp(i) - \Recall_\Base(i)$.

Before looking at ways to determine the value of $\GoodStableWeight$, note that equation (\ref{missing_weight_formula}) is useful to rule out implausible values of $\Recall_\Base(i)$ because of the constraint $\MissingWeight \geq 0$. In particular:
\begin{align}
&\frac{\BadSplitWeight + \GoodStableWeight}{\Recall_\Base(i)} \nonumber \\
& - \BadSplitWeight - \GoodStableWeight - \GoodMergeWeight \geq 0 \\
&\frac{\BadSplitWeight + \GoodStableWeight}{\Recall_\Base(i)} \geq \nonumber \\
& \BadSplitWeight + \GoodStableWeight + \GoodMergeWeight \\
& \BadSplitWeight + \GoodStableWeight \geq \nonumber \\
&\Recall_\Base(i) \cdot \left( \BadSplitWeight + \GoodStableWeight + \GoodMergeWeight \right) \\
&\Recall_\Base(i) \leq \frac{\BadSplitWeight + \GoodStableWeight}{\BadSplitWeight + \GoodStableWeight + \GoodMergeWeight}
\end{align}
The last step holds because all the annotated weights in Figure~\ref{base_exp_ideal_cluster_diagram} are non-negative, and $\GoodStableWeight$ must be positive because it includes at least $\weight(i)$.

\subsubsection{The value of $\GoodStableWeight$ according to Variant 1}

The first variant uses $\DeltaPrecision(i)$ to determine $\GoodStableWeight$, but as we will see, it works only if $\SplitRate(i) \neq \MergeRate(i)$.
\begin{align}
\DeltaPrecision(i) = &~\Precision_\Exp(i) - \Precision_\Base(i) \\
= &~\frac{\GoodMergeWeight + \GoodStableWeight}{\weight(\Exp(i))} \nonumber \\
&~ - \frac{\BadSplitWeight + \GoodStableWeight}{\weight(\Base(i))} \\
= &~\GoodMergeRate(i) + \frac{\GoodStableWeight}{\weight(\Exp(i))} \nonumber \\
&~ - \BadSplitRate(i) - \frac{\GoodStableWeight}{\weight(\Base(i))} \\
= &~ \GoodMergeRate(i) - \BadSplitRate(i) \nonumber \\
&~ + \GoodStableWeight \cdot \left( \frac{1}{\weight(\Exp(i))} - \frac{1}{\weight(\Base(i))} \right) \label{delta_precision_and_good_stable_weight}
\end{align}

So we can determine $\GoodStableWeight$ from $\DeltaPrecision(i)$ if $\weight(\Base(i)) \neq \weight(\Exp(i))$. That inequality is equivalent to $\SplitRate(i) \neq \MergeRate(i)$, because
\begin{align}
\weight(\Base(i)) = \weight(\Exp(i)) \Longleftrightarrow \SplitRate(i) = \MergeRate(i)
\end{align}
\begin{proof}
Notice that $\BadStableWeight + \GoodStableWeight > 0$,  and $\weight(\Base(i)) > 0$   and  $\weight(\Exp(i)) > 0$. We have:
\begin{align*}
&~\weight(\Base(i)) = \weight(\Exp(i)) \\
\Longleftrightarrow &~ \frac{\BadStableWeight + \GoodStableWeight}{\weight(\Base(i))} =  \frac{\BadStableWeight + \GoodStableWeight}{\weight(\Exp(i))} \\
\Longleftrightarrow &~  1 - \SplitRate(i)  =  1 - \MergeRate(i) \\
\Longleftrightarrow &~  \SplitRate(i) = \MergeRate(i)
\end{align*}
\end{proof}

Aside: when $\SplitRate(i) = \MergeRate(i)$, then equation (\ref{delta_precision_and_good_stable_weight}) says
$$\DeltaPrecision(i) = \GoodMergeRate(i) - \BadSplitRate(i)$$
and hence $\DeltaPrecision(i)$ cannot tell us anything about $\GoodStableWeight$. When that happens, Variant 2 described in the next section can help.

Finally, we can simplify equation (\ref{delta_precision_and_good_stable_weight}) a little. Define
\begin{align}
\StableWeight &= \GoodStableWeight + \BadStableWeight \\
&= (1 - \SplitRate(i)) \cdot \weight(\Base(i))
\end{align}
Then we have:
\begin{align}
\frac{1}{\weight(\Exp(i))} - \frac{1}{\weight(\Base(i))} = \frac{\SplitRate(i) - \MergeRate(i)}{\StableWeight}
\end{align}
\begin{proof}
\begin{align*}
&~ \frac{1}{\weight(\Exp(i))} - \frac{1}{\weight(\Base(i))} \\
= &~ \left( \frac{\StableWeight}{\weight(\Exp(i))} - \frac{\StableWeight}{\weight(\Base(i))} \right) /~\StableWeight \\
= &~ \left( \left(1 - \frac{\StableWeight}{\weight(\Base(i))} \right) - \left( 1 - \frac{\StableWeight}{\weight(\Exp(i))} \right) \right) /~\StableWeight \\
= &~ \left( \SplitRate(i) - \MergeRate(i) \right) /~\StableWeight
\end{align*}
\end{proof}
So equation (\ref{delta_precision_and_good_stable_weight}) says that, if $\SplitRate(i) \neq \MergeRate(i)$, then:
\begin{align}
\GoodStableWeight =&~ \left( \DeltaPrecision(i) - \GoodMergeRate(i) + \BadSplitRate(i) \right) \nonumber \\
&~ \cdot \frac{(1 - \SplitRate(i)) \cdot \weight(\Base(i))}{\SplitRate(i) - \MergeRate(i)}
\end{align}

\subsubsection{The value of $\GoodStableWeight$ according to Variant 2}

This variant does not use $\DeltaPrecision(i)$. Instead, it describes $\GoodStableWeight$ as a function of $\Precision_\Base(i)$. It can be used in addition to Variant 1, or in cases when Variant 1 cannot be applied because $\SplitRate(i) = \MergeRate(i)$, or when the value of $\DeltaPrecision(i)$ is not readily available\footnote{We will shortly use $\DeltaPrecision(T)$ for the approximate reasoning on the overall level. Users of ABCDE are not obliged to obtain an estimate of $\DeltaPrecision(T)$, but it is highly recommended for applications where cluster homogeneity is important.}.
\begin{align}
\Precision_\Base(i) = \frac{\BadSplitWeight + \GoodStableWeight}{\weight(\Base(i))}
\end{align}
Hence:
\begin{align}
\GoodStableWeight =&~ \Precision_\Base(i) \cdot \weight(\Base(i)) - \BadSplitWeight \\
=&~ \Precision_\Base(i) \cdot \weight(\Base(i)) \nonumber \\
&~ - \BadSplitRate(i) \cdot \weight(\Base(i)) \\
=&~ \weight(\Base(i)) \cdot \left( \Precision_\Base(i) - \BadSplitRate(i) \right)
\end{align}

All areas in the Venn diagram of Figure~\ref{base_exp_ideal_cluster_diagram} must have non-negative weight values, and $\GoodStableWeight$ should be at least $\weight(i)$. So for the given value of $\Precision_\Base(i)$ to make sense, it must satisfy:
\begin{align}
\frac{\weight(i)}{\weight(\Base(i))} + \BadSplitRate(i) \leq \Precision_\Base(i) \leq  1 - \GoodSplitRate(i)
\end{align}

Once we know $\GoodStableWeight$, we can easily determine $\Precision_\Exp(i)$ and hence $\DeltaPrecision(i) = \Precision_\Exp(i) - \Precision_\Base(i)$.

\subsection{Approximate reasoning for overall $\DeltaRecall$}\label{approximate_reasoning_for_overall_delta_recall}


The reasoning uses ABCDE's overall quality metrics as input, namely the $\GoodSplitRate(T)$, $\BadSplitRate(T)$, $\GoodMergeRate(T)$, $\BadMergeRate(T)$, which we will henceforth abbreviate with $\GoodBadSplitMergeRate(T)$, and, if available, $\DeltaPrecision(T)$. These metrics can be measured directly and efficiently via sampling, human judgement, and estimation~\cite{vanstadengrubb2024abcde}.

The $\GoodBadSplitMergeRate(T)$ and $\DeltaPrecision(T)$ metrics produced by ABCDE are for the whole population of items $T$, i.e. all items that are affected or unaffected by the clustering change. An item is called affected by the change iff its $\Base$ and $\Exp$ clusters do not have identical members:
\begin{align}
\AffectedItems &= \{i \in T | \Base(i) \neq \Exp(i)\} \\
\UnaffectedItems &= \{i \in T | \Base(i) = \Exp(i)\}
\end{align}

As discussed in~\cite{vanstadengrubb2024abcde}, the ABCDE quality metrics are all zero for unaffected items, and because the aggregate metrics are weighted averages, it is easy to obtain the corresponding metrics for the sub-population of affected items:
\begin{align}
\AffectedWeightFraction &= \frac{\weight(\AffectedItems)}{\weight(T)} \\
\GoodSplitRate(\AffectedItems) &= \frac{\GoodSplitRate(T)}{\AffectedWeightFraction} \\
\BadSplitRate(\AffectedItems) &= \frac{\BadSplitRate(T)}{\AffectedWeightFraction} \\
\GoodMergeRate(\AffectedItems) &= \frac{\GoodMergeRate(T)}{\AffectedWeightFraction} \\
\BadMergeRate(\AffectedItems) &= \frac{\BadMergeRate(T)}{\AffectedWeightFraction} \\
\DeltaPrecision(\AffectedItems) &= \frac{\DeltaPrecision(T)}{\AffectedWeightFraction}
\end{align}

Note that $\AffectedWeightFraction$ does not depend on sampling or human judgements; it can be computed exactly. Up to here all metrics are exact, so $\GoodSplitRate(\AffectedItems)$ is indeed the good split rate of the sub-population of affected items, etc. Division by zero here means that $\AffectedWeightFraction = 0$, in which case the clustering change is a noop and hence uninteresting from the perspective of quality assessment.

Now for the approximate part. We first create a single diagram similar to Figure~\ref{base_exp_ideal_cluster_diagram} to describe the ``expected situation'' for a ``typical'' affected item (details of this will follow). Then we reason about the ``typical'' affected item using the formulas of the previous section. We obtain a formula for $\DeltaRecall(\AffectedItems)$ as a function of $\Recall_\Base(\AffectedItems)$ that involves the quality metrics $\GoodBadSplitMergeRate(\AffectedItems)$. The formula based on Variant~1 of the previous section also uses $\DeltaPrecision(\AffectedItems)$. The formula based on Variant~2 expresses $\DeltaRecall(\AffectedItems)$ as a function of an assumed $\Precision_\Base(\AffectedItems)$ and $\Recall_\Base(\AffectedItems)$.

Finally, we can transform the approximate value of $\DeltaRecall(\AffectedItems)$ into an approximate value for overall $\DeltaRecall$:
\begin{align}
\DeltaRecall(T) = \DeltaRecall(\AffectedItems) \cdot  \AffectedWeightFraction
\end{align}

The only missing piece is how to create the diagram, which we describe next.

Firstly, we choose a positive weight for the diagram's baseline cluster $B$. The value does not really matter because the ABCDE metrics all involve ratios between areas of the diagram, so we can use for example $\weight(B) = 1$, but for clarity we do not commit to a particular choice in the formulas below.

Next, we assign a weight to the diagram's experiment cluster $E$ by reasoning as follows. The part of $B$ that is split off has weight
$$\weight(B) \cdot \SplitRate(\AffectedItems)$$
The stable part has weight
\begin{align}
\StableWeight = \weight(B) \cdot (1 - \SplitRate(\AffectedItems))
\end{align}
By merging in weight $M$, we know that
$$\weight(E) = \weight(B) \cdot (1 - \SplitRate(\AffectedItems)) + M$$
and hence
$$\MergeRate(\AffectedItems) = \frac{M}{\weight(B) \cdot (1 - \SplitRate(\AffectedItems)) + M}$$
We next solve for $M$:
\begin{align*}
&~ M = \MergeRate(\AffectedItems) \cdot \left( \weight(B) \cdot (1 - \SplitRate(\AffectedItems)) + M \right) \\
\Longleftrightarrow &~ M (1 - \MergeRate(\AffectedItems)) = \\
&~ \MergeRate(\AffectedItems) \cdot \weight(B) \cdot (1 - \SplitRate(\AffectedItems)) \\
\Longleftrightarrow &~ M = \frac{\MergeRate(\AffectedItems) \cdot \weight(B) \cdot (1 - \SplitRate(\AffectedItems))}{1 - \MergeRate(\AffectedItems)}
\end{align*}
So we have:
\begin{align}
\weight(E) =&~ \weight(B) \cdot (1 - \SplitRate(\AffectedItems)) \nonumber \\
&~ + \frac{\MergeRate(\AffectedItems) \cdot \weight(B) \cdot (1 - \SplitRate(\AffectedItems))}{1 - \MergeRate(\AffectedItems)}
\end{align}

Next, we specify weight values for the diagram's good and bad splits and merges:
\begin{align}
\GoodSplitWeight &= \GoodSplitRate(\AffectedItems) \cdot \weight(B) \\
\BadSplitWeight &= \BadSplitRate(\AffectedItems) \cdot \weight(B) \\
\GoodMergeWeight &= \GoodMergeRate(\AffectedItems) \cdot \weight(E) \\
\BadMergeWeight &= \BadMergeRate(\AffectedItems) \cdot \weight(E)
\end{align}

On the level of individual items, we always had $\weight(i) \leq \GoodStableWeight$. To mirror that for affected items, we use the pointwise metric
\begin{align}
\mathit{WeightFractionOfBaseCluster}(i) = \frac{weight(i)}{\weight(\Base(i))}
\end{align}
which is lifted to sets of items in the usual way, namely with expected values, so in particular we have:
\begin{align}
& \mathit{WeightFractionOfBaseCluster}(\AffectedItems) = \nonumber \\
&~~ \frac{\sum_{i \in \AffectedItems} weight(i) \cdot \mathit{WeightFractionOfBaseCluster}(i)}{\weight(\AffectedItems)}
\end{align}
Then in a moment we will impose the analogous constraint with
\begin{align}
\mathrm{MinGoodStableWeight} = \mathit{WeightFractionOfBaseCluster}(\AffectedItems) \cdot \weight(B)
\end{align}

\noindent \textbf{Variant 1}

\noindent If $\SplitRate(\AffectedItems) \neq \MergeRate(\AffectedItems)$, then we can take inspiration from Variant 1 and define:
\begin{align}
\mathit{GSW} =&~ ( \DeltaPrecision(\AffectedItems) - \GoodMergeRate(\AffectedItems) \nonumber \\ 
&~~ + \BadSplitRate(\AffectedItems) ) \nonumber \\
&~ \cdot \frac{(1 - \SplitRate(\AffectedItems)) \cdot \weight(B)}{\SplitRate(\AffectedItems) - \MergeRate(\AffectedItems)}
\end{align}
\qed

\noindent \textbf{Variant 2}

\noindent For a given value of $\Precision_\Base(\AffectedItems)$ that satisfies:
\begin{align}
\Precision_\Base(\AffectedItems) &\geq \mathit{WeightFractionOfBaseCluster}(\AffectedItems) \nonumber \\
&~~~~~ + \BadSplitRate(\AffectedItems) \\
\Precision_\Base(\AffectedItems) &\leq 1 - \GoodSplitRate(\AffectedItems) \\
\end{align}
We define
\begin{align}
\mathit{GSW} =&~ \weight(B) \cdot \left( \Precision_\Base(\AffectedItems) - \BadSplitRate(\AffectedItems) \right)
\end{align}
\qed

Irrespective of the variant, we can define:
\begin{align}
\GoodStableWeight =&~ \text{clip $\mathit{GSW}$ to the range $[\mathrm{MinGoodStableWeight}, \StableWeight]$}
\end{align}
Another possibility is to abandon the reasoning if $\mathit{GSW}$ is out of range.

Once $\GoodStableWeight$ is known, we also know $\BadStableWeight$:
\begin{align}
\BadStableWeight = \StableWeight - \GoodStableWeight
\end{align}

For a given positive value of $\Recall_\Base(\AffectedItems)$ that satisfies:
\begin{align}
\Recall_\Base(\AffectedItems) \leq \frac{\BadSplitWeight + \GoodStableWeight}{\BadSplitWeight + \GoodStableWeight + \GoodMergeWeight}
\end{align}
we define
\begin{align}
\mathit{Missing} = &~\frac{\BadSplitWeight + \GoodStableWeight}{\Recall_\Base(\AffectedItems)} \nonumber \\
&~ - \BadSplitWeight - \GoodStableWeight - \GoodMergeWeight
\end{align}
and finally
\begin{align}
\MissingWeight = &~\max(\mathit{Missing}, 0)
\end{align}
Another possibility is to abandon the reasoning if $\textit{Missing} < 0$.

The definition of the diagram is now complete.

From the diagram, we can obtain an approximation of $\Recall_\Exp(\AffectedItems)$, and also $\Precision_\Exp(\AffectedItems)$ if we want (it can be useful for Variant 2):
\begin{align}
&\Recall_\Exp(\AffectedItems) \approx \nonumber \\
&~~\frac{\GoodStableWeight + \GoodMergeWeight}{\GoodStableWeight + \GoodMergeWeight + \BadSplitWeight + \MissingWeight} \\
&\Precision_\Exp(\AffectedItems) \approx \frac{\GoodStableWeight + \GoodMergeWeight}{\weight(E)}
\end{align}
and use them in the following equations to obtain approximations of $\DeltaRecall(T)$ and $\DeltaPrecision(T)$:
\begin{align}
&\DeltaRecall(\AffectedItems) = \Recall_\Exp(\AffectedItems) - \Recall_\Base(\AffectedItems) \\
&\DeltaPrecision(\AffectedItems) = \Precision_\Exp(\AffectedItems) - \Precision_\Base(\AffectedItems) \\
&\DeltaRecall(T) = \DeltaRecall(\AffectedItems) \cdot  \AffectedWeightFraction \\
&\DeltaPrecision(T) = \DeltaPrecision(\AffectedItems) \cdot \AffectedWeightFraction
\end{align}

\subsection{Discussion}

The reasoning is parametric in the value of $\Recall_\Base(\AffectedItems)$, and in the case of Variant~2, also $\Precision_\Base(\AffectedItems)$. While these quantities are unknown in practice, they have an intuitive meaning. In that sense they seem superior to some alternatives that can also do the job in principle, for example the expected $\frac{\GoodStableWeight}{\StableWeight}$ of the affected items.

Evaluation with respect to ground truth clusterings can help to provide ballpark values for the parameter(s), or we can visualize $\DeltaRecall(T)$ as a graph without committing to parameter values, or we can assume specific parameter values in the absence of more information.

Graphs can provide comprehensive information. For example, for Variant~1 we can plot a graph that shows $\DeltaRecall(T)$ as a function of $\Recall_\Base(\AffectedItems)$, and for Variant~2 we can create a heatmap that shows $\DeltaRecall(T)$ as a function of $\Precision_\Base(\AffectedItems)$ and $\Recall_\Base(\AffectedItems)$. Variant~2 can likewise have a heatmap that presents $\DeltaPrecision(T)$ as a function of $\Precision_\Base(\AffectedItems)$ and $\Recall_\Base(\AffectedItems)$.

It might also make sense to pick particular values of $\Recall_\Base(\AffectedItems)$ (and $\Precision_\Base(\AffectedItems)$) and to show a single value for $\DeltaRecall(T)$ (and $\DeltaPrecision(T)$). For example, we can assume $\Recall_\Base(\AffectedItems) = 70\%$ and present the corresponding value of $\DeltaRecall(T)$. Or we can let the value depend on whether the change is recall-positive or recall-negative: if the change is recall-positive, i.e.
$$\BadSplitWeight < \GoodMergeWeight$$
then assume $\Recall_\Base(\AffectedItems) = 60\%$, and otherwise assume it is $80\%$. Doing so has a dampening effect: compared to using $70\%$ across the board, it errs on the safe side by reducing an overall recall gain and amplifying an overall recall loss.

In our practical application, we found that the back-of-the-envelope estimate of $\DeltaRecall(T)$ produced by the abovementioned 60\%/80\% technique has a high linear correlation with $\DeltaRecall$ measurements obtained by evaluation with respect to a ground truth clustering (which was constructed with recall measurements in mind as described in the introduction). It can of course be different in another application, but we found near-perfect linear correlation with a Y-intercept very close to zero, which is highly desirable. The sample data included several clustering changes that the ground truth clustering considered to be recall-positive, recall-negative and recall-neutral.

We also observed other back-of-the-envelope estimates to be plausible. For example,
\begin{align}
\mathit{JD} = \left( 1 - \frac{\StableWeight}{\weight(B) + \weight(E) - \StableWeight} \right) \AffectedWeightFraction
\end{align}
agrees very closely with the exact measurement of the $\JaccardDistance$ between $\Base$ and $\Exp$ in actual experiments.

We highly recommend performing such checks to make sure that the assumptions are reasonable for the application at hand.

\section{The $\IQ$ metric}

The ABCDE impact metrics characterize the magnitude of the clustering diff between $\Base$ and $\Exp$. It is natural to ask about the extent to which the diff contributes to the improvement of the quality of the clustering. Ideally, a large diff between $\Base$ and $\Exp$ should translate into a large improvement in the quality.

In many practical applications of clustering it is important to maintain relatively stable clusters. A large clustering diff without large quality benefits would generally be considered unacceptable, unless there are large benefits in other dimensions such as resource consumption or maintainability. Either way, it can be useful to have a metric that characterizes the relationship between the impact and the quality of a clustering change. That is exactly what the $\IQ$ metric described in this section aims to accomplish, and as we will shortly see, it can summarize the general thrust of ABCDE's zoo of metrics in a single overall number that puts the quality improvement and the magnitude of the diff in relation to each other.

The basic metrics of ABCDE tell us the exact value of the $\JaccardDistance$ between $\Base$ and $\Exp$. We focus on the $\JaccardDistance$ as the main impact metric, because it characterizes the impact in a single number and it is known to be a true distance metric for clusterings~\cite{vanstaden2024pointwise}, i.e. it provides a fundamental notion of distance that satisfies familiar mathematical properties like symmetry and the triangle inequality.

Given two clusterings $C$ and $D$, the $\JaccardDistance$ between them from the perspective of an item $i$ is
\begin{align}
\JaccardDistance_{C,D}(i) &= 1 - \frac{\weight(C(i) \cap D(i))}{\weight(C(i) \cup D(i))}
\end{align}
That is a normal pointwise metric that can be lifted to sets of items in the usual way, namely by using expected values:
\begin{align}
\JaccardDistance_{C,D}(I) &= \frac{\sum_{i \in I} \weight(i) \cdot \JaccardDistance_{C,D}(i)}{\weight(I)}
\end{align}

In particular, we know the exact value of $\JaccardDistance_{\Base, \Exp}(T)$ from the basic ABCDE impact metrics. 

For the quality aspect, we would like to know how much the $\Exp$ clustering is closer to (or further from) the $\Ideal$ clustering when compared to the $\Base$ clustering. For that, we define the pointwise metric:
\begin{align}
&\JaccardDistanceToIdealImprovement_{C,D}(i) \nonumber \\
&= \JaccardDistance_{C,\Ideal}(i) - \JaccardDistance_{D,\Ideal}(i)
\end{align}
Like all the basic ABCDE metrics, and $\DeltaRecall$, it is zero for items that were unaffected by the clustering change:
\begin{align}
C(i) = D(i) \Longrightarrow \JaccardDistanceToIdealImprovement_{C,D}(i) = 0
\end{align}
Swapping the roles of $C$ and $D$ simply negates its value:
\begin{align}
&\JaccardDistanceToIdealImprovement_{C,D}(i) \nonumber \\
&= -\JaccardDistanceToIdealImprovement_{D,C}(i)
\end{align}
A pointwise metric is lifted to sets of items by using expected values. On the lifted level, we have:
\begin{align}
&\JaccardDistanceToIdealImprovement_{C,D}(I) \nonumber \\
&= \frac{\sum_{i \in I} \weight(i) \cdot \JaccardDistanceToIdealImprovement_{C,D}(i)}{\weight(I)} \\
&= \frac{\sum_{i \in I} \weight(i) \cdot  \left( \JaccardDistance_{C,\Ideal}(i) - \JaccardDistance_{D,\Ideal}(i) \right)}{\weight(I)} \\
&= \JaccardDistance_{C,\Ideal}(I) - \JaccardDistance_{D,\Ideal}(I)
\end{align}
and of course
\begin{align}
&\JaccardDistanceToIdealImprovement_{C,D}(I) \nonumber \\
&= -\JaccardDistanceToIdealImprovement_{D,C}(I)
\end{align}

On the overall level, the $\JaccardDistanceToIdealImprovement$ metric has a simple interpretation:
\begin{itemize}
\item $\JaccardDistanceToIdealImprovement_{\Base,\Exp}(T) < 0$ means the clustering change is quality-negative. The magnitude says how quality-negative the change is.
\item $\JaccardDistanceToIdealImprovement_{\Base,\Exp}(T) > 0$ means the clustering change is quality-positive. The magnitude says how quality-positive the change is.
\item $\JaccardDistanceToIdealImprovement_{\Base,\Exp}(T) = 0$ means the clustering change is quality-neutral.
\end{itemize}

The $\IQ$ metric simply puts the overall quality improvement in relation to the overall magnitude of the diff:
\begin{align}
\IQ_{\Base,\Exp}(T) = \frac{\JaccardDistanceToIdealImprovement_{\Base,\Exp}(T)}{\JaccardDistance_{\Base,\Exp}(T)}
\end{align}
The case where the denominator is zero is totally uninteresting from an evaluation point of view: $\Base$ and $\Exp$ are then identical. That follows from the ``identity of indiscernibles'' property of a true distance metric.

Before looking at the interpretation of $\IQ$, we quickly remark that swapping the roles of $\Base$ and $\Exp$ simply negates its value:
\begin{align}
\IQ_{\Base, \Exp}(T) = -\IQ_{\Exp, \Base}(T)
\end{align}

The rest of this section will abbreviate $\IQ_{\Base, \Exp}(T)$ as $\IQ$.

$\IQ$ can take on values in the range $[-1, 1]$. The triangle inequality prevents it from taking on other values:
\begin{itemize}
\item From \\
$\JaccardDistance_{\Base,\Ideal} \leq \JaccardDistance_{\Base,\Exp} + \JaccardDistance_{\Exp,\Ideal}$ \\
we know \\
$\JaccardDistance_{\Base,\Ideal} - \JaccardDistance_{\Exp,\Ideal} \leq \JaccardDistance_{\Base,\Exp}$ \\
Dividing by $\JaccardDistance_{\Base,\Exp}$ yields $\IQ \leq 1$.
\item From \\
$\JaccardDistance_{\Exp,\Ideal} \leq \JaccardDistance_{\Exp,\Base} + \JaccardDistance_{\Base,\Ideal}$ \\
and symmetry we know \\
$-\JaccardDistance_{\Base,\Exp} \leq \JaccardDistance_{\Base,\Ideal} - \JaccardDistance_{\Exp,\Ideal}$ \\
Dividing by $\JaccardDistance_{\Base,\Exp}$ yields $-1 \leq \IQ$.
\end{itemize}
It is relatively easy to interpret $\IQ$ values: $IQ = x\%$ means for every 100 units (steps) $\Exp$ is away from $\Base$, it is $x$ units closer to the $\Ideal$ clustering. So, in particular,
\begin{itemize}
\item $\IQ = 1$, i.e. $100\%$, is the maximal value of $\IQ$. It means that \\
$\JaccardDistanceToIdealImprovement_{\Base,\Exp}(T) = \JaccardDistance_{\Base,\Exp}(T)$. So $\Exp$ is somewhere on the shortest path from $\Base$ to $\Ideal$.
\item $\IQ > 0$ means $\Exp$ is a quality improvement.
\item $\IQ = 0$, i.e. $0\%$, means $\Exp$ is quality-neutral.
\item $\IQ < 0$ means $\Exp$ is a quality regression.
\item $\IQ = -1$, i.e. $-100\%$, is the minimal value of $\IQ$. It means that $\JaccardDistanceToIdealImprovement_{\Base,\Exp}(T)$ is negative and equal in magnitude to $\JaccardDistance_{\Base,\Exp}(T)$. So $\Base$ is somewhere on the shortest path from $\Exp$ to $\Ideal$.
\end{itemize}

\begin{figure*}[!]
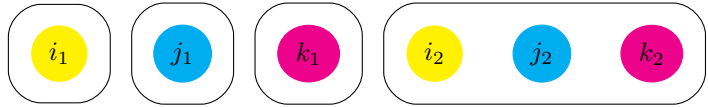
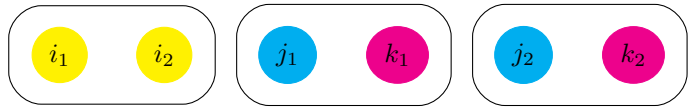
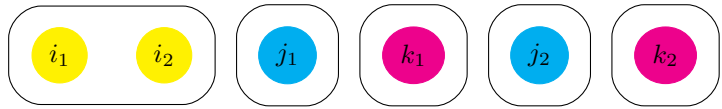
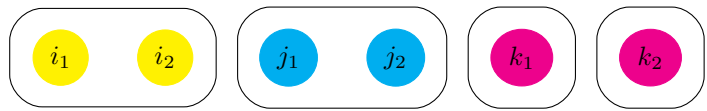
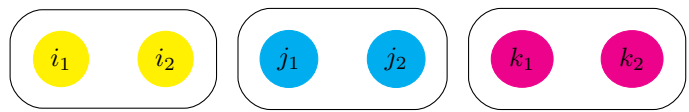

    \centering
    \begin{subfigure}[t]{\textwidth}
        \centering
        \begin{dot2tex}[dot,mathmode]
            graph {
        		subgraph cluster_1 {
        		    style=rounded
        			node [color=yellow, style=filled]
        			i2 [label=i_2]
        			i1 [label=i_1]
        		}
        		subgraph cluster_2 {
        		    style=rounded
        			node [color=cyan, style=filled]
        			j2 [label=j_2]
        			j1 [label=j_1]
        		}
        		subgraph cluster_3 {
        		    style=rounded
        		    node [color=magenta, style=filled]
        			k2 [label=k_2]
        			k1 [label=k_1]
        		}
        	}
        \end{dot2tex}
        \caption{The $\Ideal$ clustering.}
    \end{subfigure}
    ~
    \par\bigskip
    \begin{subfigure}[t]{\textwidth}
        \centering
        \begin{dot2tex}[dot,mathmode]
            graph {
        		subgraph cluster_1 {
        		    style=rounded
        			node [color=magenta, style=filled]
        			k1 [label=k_1]
        			node [color=cyan, style=filled]
        			j1 [label=j_1]
        			node [color=yellow, style=filled]
        			i1 [label=i_1]
        		}
        		subgraph cluster_2 {
        		    style=rounded
        			node [color=magenta, style=filled]
        			k2 [label=k_2]
        			node [color=cyan, style=filled]
        			j2 [label=j_2]
        			node [color=yellow, style=filled]
        			i2 [label=i_2]
        		}
        	}
        \end{dot2tex}
        \caption{The $\Base$ clustering.}
    \end{subfigure}
    ~
    \par\bigskip
    \begin{subfigure}[t]{\textwidth}
        \centering
        \begin{dot2tex}[dot,mathmode]
            graph {
        		subgraph cluster_1 {
        		    style=rounded
        			node [color=cyan, style=filled]
        			j1 [label=j_1]
        			node [color=yellow, style=filled]
        			i1 [label=i_1]
        		}
        		subgraph cluster_2 {
        		    style=rounded
        			node [color=magenta, style=filled]
        			k1 [label=k_1]
        		}
        		subgraph cluster_3 {
        		    style=rounded
        			node [color=magenta, style=filled]
        			k2 [label=k_2]
        			node [color=cyan, style=filled]
        			j2 [label=j_2]
        			node [color=yellow, style=filled]
        			i2 [label=i_2]
        		}
        	}
        \end{dot2tex}
        \caption{An $\Exp$ clustering with $\IQ = 31.25\%$.}
    \end{subfigure}
    ~
    \par\bigskip
    \begin{subfigure}[t]{\textwidth}
        \centering
        \begin{dot2tex}[dot,mathmode]
            graph {
        		subgraph cluster_1 {
        		    style=rounded
        			node [color=yellow, style=filled]
        			i1 [label=i_1]
        		}
        		subgraph cluster_2 {
        		    style=rounded
        			node [color=cyan, style=filled]
        			j1 [label=j_1]
        		}
        		subgraph cluster_3 {
        		    style=rounded
        			node [color=magenta, style=filled]
        			k1 [label=k_1]
        		}
        		subgraph cluster_4 {
        		    style=rounded
        			node [color=magenta, style=filled]
        			k2 [label=k_2]
        			node [color=cyan, style=filled]
        			j2 [label=j_2]
        			node [color=yellow, style=filled]
        			i2 [label=i_2]
        		}
        	}
        \end{dot2tex}
        \caption{An $\Exp$ clustering with $\IQ = 37.50\%$.}
    \end{subfigure}
    ~
    \par\bigskip
    \begin{subfigure}[t]{\textwidth}
        \centering
        \begin{dot2tex}[dot,mathmode]
            graph {
        		subgraph cluster_1 {
        		    style=rounded
        			node [color=yellow, style=filled]
        			i2 [label=i_2]
        			i1 [label=i_1]
        		}
        		subgraph cluster_2 {
        		    style=rounded
        		    node [color=magenta, style=filled]
        			k1 [label=k_1]
        			node [color=cyan, style=filled]
        			j1 [label=j_1]
        		}
        		subgraph cluster_4 {
        		    style=rounded
        			node [color=magenta, style=filled]
        			k2 [label=k_2]
        			node [color=cyan, style=filled]
        			j2 [label=j_2]
        		}
        	}
        \end{dot2tex}
        \caption{An $\Exp$ clustering with $\IQ = 64.71\%$.}
    \end{subfigure}
    ~
    \par\bigskip
    \begin{subfigure}[t]{\textwidth}
        \centering
        \begin{dot2tex}[dot,mathmode]
            graph {
        		subgraph cluster_1 {
        		    style=rounded
        			node [color=yellow, style=filled]
        			i2 [label=i_2]
        			i1 [label=i_1]
        		}
        		subgraph cluster_2 {
        		    style=rounded
        			node [color=cyan, style=filled]
        			j1 [label=j_1]
        		}
        		subgraph cluster_3 {
        		    style=rounded
        		    node [color=magenta, style=filled]
        			k1 [label=k_1]
        		}
        		subgraph cluster_4 {
        		    style=rounded
        			node [color=cyan, style=filled]
        			j2 [label=j_2]
        		}
        		subgraph cluster_5 {
        		    style=rounded
        			node [color=magenta, style=filled]
        			k2 [label=k_2]
        		}
        	}
        \end{dot2tex}
        \caption{An $\Exp$ clustering with $\IQ = 60.00\%$.}
    \end{subfigure}
    ~
    \par\bigskip
    \begin{subfigure}[t]{\textwidth}
        \centering
        \begin{dot2tex}[dot,mathmode]
            graph {
        		subgraph cluster_1 {
        		    style=rounded
        			node [color=yellow, style=filled]
        			i2 [label=i_2]
        			i1 [label=i_1]
        		}
        		subgraph cluster_2 {
        		    style=rounded
        			node [color=cyan, style=filled]
        			j2 [label=j_2]
        			j1 [label=j_1]
        		}
        		subgraph cluster_3 {
        		    style=rounded
        		    node [color=magenta, style=filled]
        			k1 [label=k_1]
        		}
        		subgraph cluster_4 {
        		    style=rounded
        			node [color=magenta, style=filled]
        			k2 [label=k_2]
        		}
        	}
        \end{dot2tex}
        \caption{An $\Exp$ clustering with $\IQ = 80.77\%$.}
    \end{subfigure}
    ~
    \par\bigskip
    \begin{subfigure}[t]{\textwidth}
        \centering
        \begin{dot2tex}[dot,mathmode]
            graph {
        		subgraph cluster_1 {
        		    style=rounded
        			node [color=yellow, style=filled]
        			i2 [label=i_2]
        			i1 [label=i_1]
        		}
        		subgraph cluster_2 {
        		    style=rounded
        			node [color=cyan, style=filled]
        			j2 [label=j_2]
        			j1 [label=j_1]
        		}
        		subgraph cluster_3 {
        		    style=rounded
        		    node [color=magenta, style=filled]
        		    k2 [label=k_2]
        			k1 [label=k_1]
        		}
        	}
        \end{dot2tex}
        \caption{An $\Exp$ clustering with $\IQ = 100.00\%$.}
    \end{subfigure}
    \caption{Several $\Exp$ clusterings that have only quality improvements, i.e. only good splits and good merges and no bad splits or bad merges, and their associated $\IQ$ values. All items are assumed to have equal weights.}\label{iq_for_clustering_improvements}
\end{figure*}

Most clustering changes that aim to improve the quality will have $\IQ < 1$, which means the $\Exp$ clustering will not lie somewhere on the shortest path from $\Base$ to $\Ideal$. It is unrealistic for the diff to contain only improvements. Even if there are no bad splits or bad merges, $\Exp$ is still not guaranteed to be on the shortest path to $\Ideal$: as the examples in Figure~\ref{iq_for_clustering_improvements} show, there are many such changes that have a much bigger $\JaccardDistance$ between $\Base$ and $\Exp$ than the improvement in distance to the $\Ideal$ clustering.

\subsection{Geometric interpretation of $\IQ$}

\begin{figure}[t!]
\makebox[\linewidth][c]{%
\begin{subfigure}[b]{1.4\textwidth}
\centering
\includegraphics[width=\textwidth]{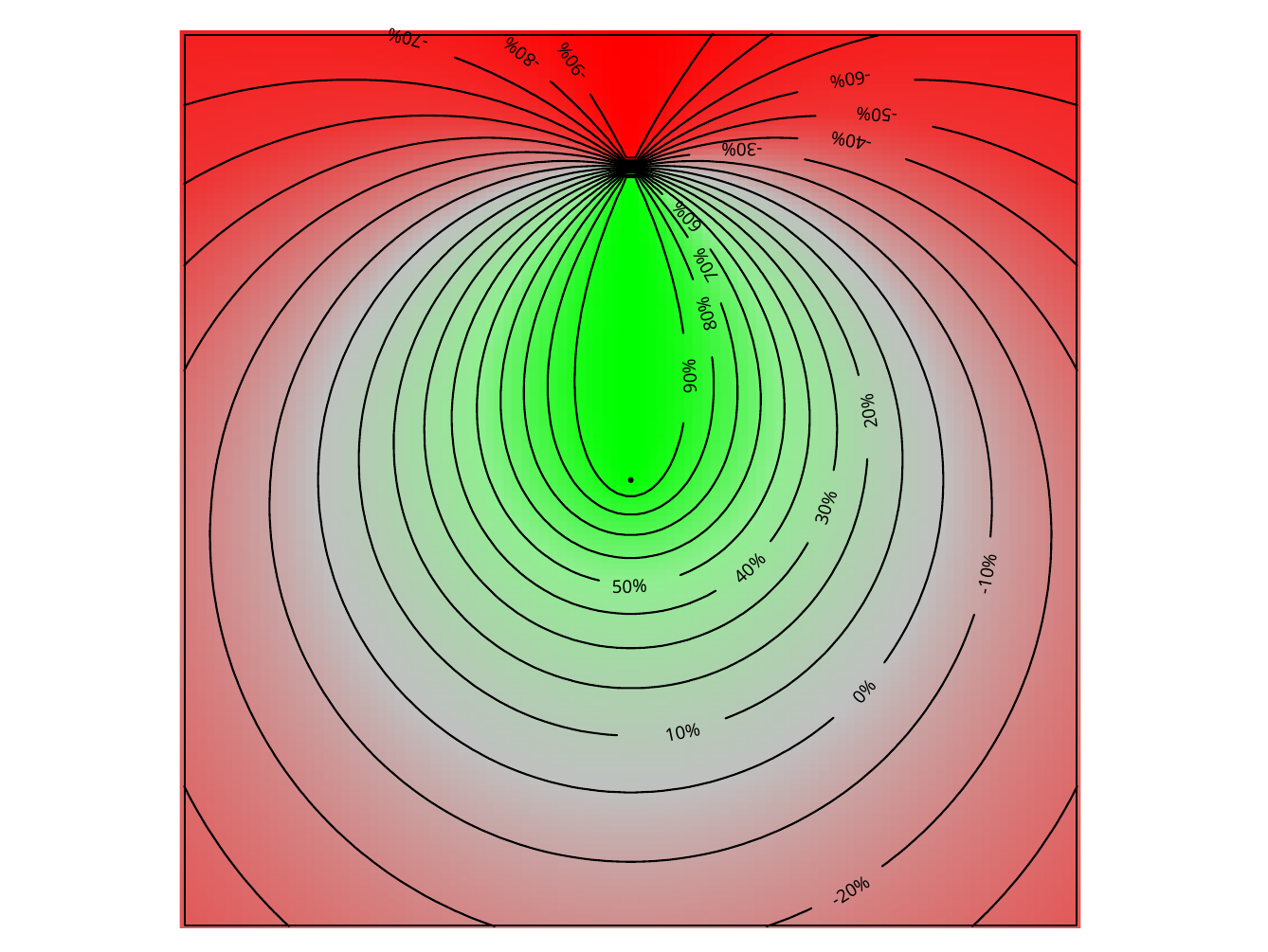}
\end{subfigure}%
}
\caption{A graphical representation of the $\IQ$ metric in 2D space. The ideal clustering is the dot at the center, while $\Base$ is some distance above it at the point where the curves meet. Each value of $\IQ$ specifies a curve on which $\Exp$ is located somewhere. Quality-positive changes have $\IQ > 0$ and are colored green. Quality-neutral changes have $\IQ = 0$ and are colored grey. Quality-negative changes have $\IQ < 0$ and are colored red.}\label{iq_onion}
\end{figure}

Humans have a good intuition about Euclidean space, which is the fundamental space of geometry and intended to represent the physical space as we perceive it in our everyday lives. It is much easier for us to think about Euclidean space and Euclidean distance than it is to think about the space of clusterings and the $\JaccardDistance$. To gain an intuitive understanding of some metric in clustering space, such as $\IQ$, it can be fruitful to consider its counterpart in Euclidean space.

2D Euclidean space, i.e. $\mathbb{R}^2$, is the easiest to visualize. Position the $\Ideal$ clustering at the origin, and position the $\Base$ clustering some arbitrary distance $d$ above the origin on the Y-axis (we do not know the exact distance, but that does not matter, because the resulting figure is scale-invariant). For each coordinate $(x, y)$, we can suppose that $\Exp$ is at $(x, y)$, and we calculate the corresponding value of $f(x, y, d)$, which is defined as the $\IQ$ of $\Exp$ with respect to the $\Base$ and the $\Ideal$ coordinates using the familiar Euclidean distance:
\begin{align*}
f(x, y, d) &= \IQ_{\Base,\Exp} \\
&= \frac{\JaccardDistance_{\Base,\Ideal} - \JaccardDistance_{\Exp,\Ideal}}{\JaccardDistance_{\Base,\Exp}} \\
&= \frac{d - \sqrt{x^2 + y^2}}{\sqrt{x^2 + (d - y)^2}}
\end{align*}
Plotting $f$ for a positive constant value of $d$ yields Figure~\ref{iq_onion}.

A given value of $\IQ$ traces out a curve, which gives the possible positions of an $\Exp$ clustering with that $\IQ$ metric relative to the $\Base$ and the $\Ideal$ clusterings. The curve of $\IQ = 0$ is a perfect circle. Since $\IQ_{\Base,\Exp}(T) = -\IQ_{\Exp,\Base}(T)$, the parts of the diagram inside the circle and outside the circle are isomorphic. We can exploit that to visualize a negative value of $\IQ$ more conveniently: negate the value of $\IQ$ and switch the roles of $\Base$ and $\Exp$ in the diagram.

Our intuitive understanding of the 2D visualization transfers nicely to clusterings:
\begin{itemize}
\item The space of clusterings (in which $\JaccardDistance$ operates) is very high-dimensional. \\
It is easy for us to imagine a 3D figure, very much like an onion, where the curve of a given value of $\IQ$ traces out a scale of the onion, and $\Exp$ is somewhere on the scale. So the same intuition holds in higher dimensions.
\item The space of clusterings is bounded, since $\JaccardDistance$ is always in the range $[0, 1)$, and it is also discrete.
\begin{itemize}
\item The boundedness simply means that some positions in the diagram might not be ``reachable'' by an $\Exp$ clustering. In practical cases, where $\Base$ and $\Ideal$ are fairly closely together, this should not have any effect on positive values of $\IQ$. If a curve has unreachable positions, our interpretation of it remains valid: a curve is a set of points that includes the actual position of $\Exp$.
\item The discreteness also means that some positions on the diagram might not be ``reachable'' by an $\Exp$ clustering. As discussed in the point on boundedness, this does not affect our interpretation of a curve.
\end{itemize}
\item The space of clusterings is not curved/warped in unexpected ways -- the $\JaccardDistance$ grows/shrinks proportionally to the changes in cluster membership.
\end{itemize}

\subsection{Approximate reasoning for $\IQ$}

The denominator of $\IQ$, namely $\JaccardDistance_{\Base,\Exp}(T)$, is computed exactly by ABCDE as part of the basic impact metrics.

The numerator, however, is plagued by the fact that we have very limited knowledge about the $\Ideal$ clustering in practical applications. One way to tackle that is with the approximate reasoning we developed for $\DeltaRecall(T)$. In particular, the diagram we constructed in Section~\ref{approximate_reasoning_for_overall_delta_recall} can provide the following approximations:
\begin{align}
&\JaccardDistance_{\Base,\Ideal}(\AffectedItems) \approx \nonumber \\
&~~~ 1 - \frac{\BadSplitWeight + \GoodStableWeight}{\weight(B) + \GoodMergeWeight + \MissingWeight} \\
&\JaccardDistance_{\Exp,\Ideal}(\AffectedItems) \approx \nonumber \\
&~~~ 1 - \frac{\GoodStableWeight + \GoodMergeWeight}{\weight(E) + \BadSplitWeight + \MissingWeight}
\end{align}
Next, we can multiply them with $\AffectedWeightFraction$ to obtain approximations for:
\begin{align}
&\JaccardDistance_{\Base,\Ideal}(T) = \nonumber \\
&~~~\JaccardDistance_{\Base,\Ideal}(\AffectedItems) \cdot \AffectedWeightFraction \\
&\JaccardDistance_{\Exp,\Ideal}(T) = \nonumber \\
&~~~\JaccardDistance_{\Exp,\Ideal}(\AffectedItems) \cdot \AffectedWeightFraction
\end{align}
Finally, we can subtract them to get an approximation of the numerator of $\IQ$:
\begin{align}
&\JaccardDistanceToIdealImprovement_{\Base,\Exp}(T) = \nonumber \\
&~~~\JaccardDistance_{\Base,\Ideal}(T) - \JaccardDistance_{\Exp,\Ideal}(T)
\end{align}

Since this is an approximation, it might yield values for $\IQ$ that are not perfectly inside the range, but we can clip the result of the division to $[-1,1]$.

\section{The quality of a single clustering}

ABCDE can estimate the difference in quality between a $\Base$ and an $\Exp$ clustering. It reports overall $\DeltaPrecision(T)$ and $\DeltaRecall(T)$ estimates, where
\begin{align*}
\DeltaPrecision(T) &= \Precision_\Exp(T) - \Precision_\Base(T) \\
\DeltaRecall(T) &= \Recall_\Exp(T) - \Recall_\Base(T)
\end{align*}

In order to minimize the overhead of the human judgements, and to keep the confidence intervals small, ABCDE does not estimate the terms $\Precision_\Exp(T)$ and $\Precision_\Base(T)$ separately. Instead, it uses a more sophisticated technique that will, on the basis of the clustering diffs between $\Base$ and $\Exp$, sample pairs of items for human judgement, and it will use the results to estimate $\DeltaPrecision(T)$ directly.

While delta quality metrics are sufficient for individual experiments/changes, it can be very useful for developers and management to know the absolute $\Precision$ and $\Recall$ of a clustering snapshot, i.e. the quality of the currently deployed clustering in absolute terms\footnote{Note that, even if the clustering algorithm is not changed at all, the absolute $\Precision$ and $\Recall$ values will not necessarily be stable over time, because of new/deleted items, changes in item properties, changes in item weights (because of changes in popularity), etc. So the absolute metrics and the headroom must always be understood as pertaining to the algorithm and the data of a specific point in time.}. This information has several applications:
\begin{itemize}
\item Understanding the quality headroom.
\item Deciding which efforts to prioritize, e.g. efforts to improve the $\Precision$ or the $\Recall$ of the clustering.
\item Deciding on the requirements for acceptable clustering changes. For example, specifying desirable $\Precision$ and $\Recall$ tradeoffs on the basis of the current quality situation.
\end{itemize}

\subsection{Measuring the absolute quality of a single clustering}

On a regular basis, for example once per quarter or once per year, we can use ABCDE to estimate the absolute $\Precision$ of the current clustering snapshot. That can be done because there is a trivial reference clustering with perfect $\Precision$. We define the $\PerfectPrecision$ clustering to be the clustering where every item is in a cluster by itself:
\begin{align}
\PerfectPrecision = \{\{i\} | i \in T \}
\end{align}
This clustering has $\Precision = 1$, and the minimum $\Recall$ value any clustering could possibly have, which we will denote as $\MinimumRecall$. The value of $\MinimumRecall$ will be much bigger than a small positive $\epsilon$ when many clusters are legitimately small.

We can run ABCDE, with the clustering snapshot $C$ as $\Base$, and $\PerfectPrecision$ as $\Exp$, and the weights those of the clustering snapshot\footnote{Weight assignment schemes are briefly discussed in Section 3.1 of~\cite{vanstadengrubb2024abcde}. If intrinsic importance values are used as weights, then they can simply be used as-is. If item weights are based on the importance/popularity of clusters that were produced in the past, and the scheme from Section 3.1 in~\cite{vanstadengrubb2024abcde} is used where each clustering has an associated map of per-item weights, then we can simply use the per-item weights of the clustering snapshot $C$.}. That will be a fairly large clustering change, so we can send more pairs of items for human judgement than we would for normal experiments. The resulting quality metrics can be interpreted as follows:
\begin{itemize}
\item $\DeltaPrecision(T)$ shows the $\Precision$ headroom, i.e. how far the overall $\Precision$ of $C$ is from 100\%.
\begin{align}
\DeltaPrecision(T) &= \Precision_\Exp(T) - \Precision_\Base(T) \\
&= 1 - \Precision_C(T)
\end{align}
Hence we can estimate the absolute $\Precision$ of $C$:
\begin{align}
\Precision_C(T) = 1 - \DeltaPrecision(T)
\end{align}
\item $\DeltaRecall(T)$ will not be positive.
\begin{align}
\DeltaRecall(T) &= \Recall_\Exp(T) - \Recall_\Base(T) \\
&= \MinimumRecall - \Recall_C(T)
\end{align}
Hence we can estimate the absolute $\Recall$ of $C$:
\begin{align}
\Recall_C(T) = \MinimumRecall - \DeltaRecall(T)
\end{align}
We don't really know the absolute value of $\MinimumRecall$, but we will know that $C$ has an absolute overall $Recall$ that is $|\DeltaRecall(T)|$ more than $\MinimumRecall$.
\end{itemize}
So, in summary, this experiment will inform us about how far the clustering snapshot is below 100\% $\Precision$, and how far the clustering snapshot is above $\MinimumRecall$.

These are likely the best and most realistic estimates of the absolute overall $\Precision$ and $\Recall$ of a clustering snapshot we can obtain in practice.

In principle, we could run another experiment where $\Exp$ is the trivial reference clustering with perfect $\Recall$ where all the items are put together in one giant cluster:
\begin{align}
\PerfectRecall = \{T\}
\end{align}
Unfortunately this experiment does not yield much information in practice. The reason is that humans can judge relatively few pairs of items, and it is difficult to discover many/any good merges among the sampled merge pairs when there are billions of items and a huge number of ideal clusters (i.e. a huge number of true equivalence classes). As a consequence, the estimation math will be forced to conclude that the clustering snapshot's $\Recall$ is not appreciably lower than that of $\PerfectRecall$, because the human judgements did not provide evidence to the contrary.


\section{Conclusion}

This paper equips ABCDE with more quality metrics, in particular:
\begin{itemize}
\item $\DeltaRecall$. The technique described in this paper uses sophisticated back-of-the-envelope reasoning to express $\DeltaRecall$ as a function of the baseline quality of the items that experienced change. The reasoning leverages the rich information provided by the basic ABCDE metrics. It also uses some auxiliary measurements about the clusterings that are easy to compute and do not require human judgement.
\item $\IQ$. The $\IQ$ metric summarizes the general thrust of ABCDE's zoo of impact and quality metrics in a single overall number that puts the quality improvement and the magnitude of the diff in relation to each other.
\item Absolute $\Precision$ and $\Recall$ estimates for a single clustering snapshot. They do not need new techniques or modifications to ABCDE. Instead, we simply use ABCDE to evaluate the diff between the clustering snapshot and a special reference clustering.
\end{itemize}

The basic ABCDE metrics and the ones of this paper complement each other. Taken together, they provide an extensive set of metrics for characterizing and understanding clustering changes in practical applications.

\bibliographystyle{plain}
\bibliography{main}

\end{document}